\documentclass[12pt]{article}
\usepackage[T2A]{fontenc}
\usepackage[utf8]{inputenc}
\usepackage[english]{babel}

\usepackage{amsmath}
\usepackage{amsfonts}
\usepackage{amssymb}
\usepackage{graphicx}
\usepackage{wrapfig}
\usepackage{color}
\usepackage{cases}

\usepackage{multicol}

\usepackage{geometry}
\usepackage[colorlinks=true, linkcolor=blue,urlcolor=red,citecolor=green]{hyperref}

\newcommand{\comm}[1]{} 

\begin{document}

\begin{center}
\Large {\bf  Gravitational Waves in High Energy Fixed-Target Collisions}
\end{center}

\bigskip
\bigskip

\begin{center}
	D.V.\,Fursaev$^\dag$ and V.A.\,Tainov$^{\dag\ddag}$
\end{center}

\bigskip

\begin{center}
	{\dag \it  Bogoliubov Laboratory of Theoretical Physics\\
		Joint Institute for Nuclear Research\\
		141980, Dubna, Russia}\\
	\ddag   {\it Dubna State University, 141980, Dubna, Russia\\ }
	\medskip
\end{center}
\bigskip

\begin{abstract}
The gravitational field of two-body system, a high energetic particle and a massive particle at rest, is studied in the linearized Einstein gravity. 
The ultrarelativistic particle yields a plane-fronted gravitational shockwave which perturbes gravitational field of the particle at rest. The problem can be also considered as a fixed-target high energy collision. We show that this collision 
is accompanied by the gravitational radiation, as is expected from the earlier results on the high-energy scattering. The new effect is a secondary spherical gravitational shockwave when the initial shockwave hits the massive particle. In the considered approximation the flux of gravitational radiation and the amplitude of the spherical shockwave are found in an analytic form.  The suggested approach is also applicable when the null particle is replaced by plane null shells of a general profile.
Implications of these effects for astrophysics are shortly discussed.
\end{abstract}

\newpage

\section{Introduction}
\label{sec:1}

Scattering of ultrahigh energy particles attracts  persistent attention for a number of reasons. One of the motivations 
is related to the seminal observation by 't Hooft \cite{tHooft:1987vrq}  on the graviton dominance at the Planckian energies. Another reason  is a hypothetical possibility of black hole productions in certain scenario with large extra dimensions, see e.g. \cite{Eardley:2002re} and references therein.

It is well known that head-on collisions of two point-like massless particles  generate gravitational waves \cite{DEath:1976bbo}-\cite{DEath:1992nmz}. In this work we focus on  two-body problems of a different type. These are fixed-target collisions of a high energetic particle with a particle at rest.  We go to the extreme limit when the velocity of the moving particle equals the speed of light. 

The head-on collisions can be obtained from our setup by an ultra-boost of the particle at rest. Therefore some effects of the head-on collisions, such as the gravitational radiation, are expected in the considered problem. We will see that this is indeed the case.
However, the fixed-target collisions  also possess their own unique features, such as secondary shockwaves. 

In this paper we solve
the Einstein equations sourced by the two point-like bodies, a massive and a null (massless) particles. The gravitational field of the each body alone is known, and we find corrections to these fields in the linearized approximation. The gravitational field of the null particle is a plane-fronted gravitational shockwave (GSW). So we calculate  perturbations of the gravitational field of the massive particle generated by the shockwave. 

If the equation of the shockwave front trajectory is $u=0$, $u$ being a retarded time, the perturbations live in the future to the front, $u>0$
(since $u=0$ is also the past event horizon of the null particle). The perturbations are solutions to a characteristic Cauchy problem with the Cauchy data at $u=0$. Such characteristic problems for perturbations of scalar and Maxwell fields caused by the GSW have been recently studied in \cite{Fursaev:2024czx}, \cite{Fursaev:2025did}.

The Cauchy data on $u=0$ are determined solely by the gravitational memory effects \cite{Fursaev:2025wvh}, left after the action of  gravitational shockwaves on the fields, and can be expressed in a geometric way as a transformation of fields under the Penrose supertranslations \cite{Penrose:1972xrn}. A technical challenge of this work is to formulate the Cauchy problem for metric perturbations in a self-consistent way by taking into account the gravitational constraints.   

Gravitational shockwaves produced by null particles moving in background gravitational fields is not a new research subject. 
They have been studied starting from the paper by Dray and {'t Hooft} \cite{tHooft:1985NPB} in 1985 and used in different applications, including holographic descriptions of quantum theories. However, in this and subsequent publications, see e.g. \cite{Sfetsos:1994xa}, the background geometries had an essential restriction: the supertranslations there were isometries of the hypersurface $u=0$  (in fact, they are infinite-dimensional isometries \cite{Blau:2015nee}). This property holds, e.g., when the hypersurface $u=0$ is a Killing horizon. It will be shown below that, given this restriction, the null particle does not perturb the background geometry and it does not produce any radiation.  Thus, the gravitational radiation accompanying the fixed target collisions can be traced to the violation of the supertranslation symmetries.
	
The work is organized as follows.  In Section \ref{sec:2} we develop an approach to gravitational perturbations in collisions 
of a null matter and a massive object. Although our primary interest is in collisions of null and massive particles, the suggested approach can be used in a more general case of plane null shells, see  definitions in Section \ref{sec:2.1}.  
The gravitational memory effects for particle trajectories, scalar and electromagnetic fields after the motion of plane fronted GSW are described in Section \ref{sec:2.2}.  The case of the Maxwell field is instructive: it teaches one how the memory effects appear in the theories with constraints and how the supertranslations act on the Cauchy data.
The Penrose prescription \cite{Penrose:1972xrn} for shockwaves requires that two manifolds are glued together along the shockwave front.  We analyze this procedure in  Section \ref{sec:2.3} in the linearized gravity theory. We use a canonical approach to metric perturbations and their conjugated momenta at $u=0$ to find the relation between stress-energy density of a null shell and a surface curvature density at $u=0$. Here standard canonical analysis cannot be applied straightforwardly, since perturbations are considered at the degenerate hypersurface. Although definitions through metric variations are related to the Israel-Barrabes null shell formalism, they are more convenient for our purposes.  In Section \ref{sec:2.4} we formulate the characteristic Cauchy problem for metric perturbations. In accord with the memory effect the initial perturbations of the metric in the tangent space to $u=0$ are the supertranslations of the corresponding components of the background metric. We demonstrate that this choice is consistent with gravitational constraints. The initial data for the rest components are completely fixed by the surface curvature density. The subsequent calculations can be carried out in a convenient gauge. We use the de Donder gauge.

Gravitational effects in the fixed target collisions are discussed in the second half of the paper, in Section \ref{sec:3}. 
A solution for perturbations caused by a GSW from a plane null shell is given in an integral form in Section \ref{sec:3.1}. 
Late-time asymptotics of the perturbations and the flux of outgoing bulk gravitation radiation in collision of a null and massive particles are calculated in Section \ref{sec:3.2}.
The angular distribution of the intensity of the gravitational radiation  is obtained in a analytic form. The new feature of gravitational effects in fixed-target collisions is non-analyticity of the perturbations on a future null-cone of the event when the shockwave front hits the massive source. We demonstrate, in Section \ref{sec:3.4}, that this non-analyticity is a spherical shock gravitational wave. The non-analyticity appears as a delta-function singularity at the light cone of the Newman-Penrose invariant $\Psi_4$.

Concluding remarks with a review of possible applications of these results are presented in Section \ref{sec:4}. 
Definitions and some details regarding Sections \ref{sec:2.3}, \ref{sec:2.4} are presented in Appendix \ref{App1}. Asymptotic form of perturbations in Section \ref{sec:3.2}  is further discussed in Appendix \ref{App2}.

\section{The problem for perturbations}\label{sec:2}
\setcounter{equation}0
\subsection{The two-body system}\label{sec:2.1}

Consider two particles, one is massive and another is massless (null). Suppose that in the Minkowsky coordinates the massive particle is located 
at the center of coordinates, $x=y_i=0$, $i=1,2$, while the massless particle moves along the $x$ axis, from the left to the right, and has coordinates
$y_1=-a, y_2=0$. Here $a$ is an impact parameter between the two particles. 
In what follows we use the retarded  and advanced time coordinates $u=t-x$, $v=t+x$, where the Minkowsky metric is
$ds^2=- dudv+dy_i^2$.

We want to solve the Einstein equations 
\begin{equation}\label{1.0}
2G^{\mu\nu}= \kappa \left(T_{\mbox{\tiny{(m)}}}^{\mu\nu}+T_{\mbox{\tiny{(n)}}}^{\mu\nu} \right)~,~\kappa=16\pi G~, 
\end{equation}
sourced by the stress tensors of the massive and  null particles, $T_{\mbox{\tiny{(m)}}}^{\mu\nu}$ and $T_{\mbox{\tiny{(n)}}}^{\mu\nu}$, respectively. The both tensors are  distributions with supports on trajectories of particles. 

Our analysis will be restricted by the linearized theory when the Einstein tensor $G_{\mu\nu}$ is linear in metric perturbations over the Minkowsky background. If the massive particle was alone, $T_{(\mbox{\tiny{n}})}^{\mu\nu}=0$, the metric in the linearized gravity theory would be
\begin{equation}\label{1.1}
g_{\mu\nu}=\eta_{\mu\nu}+\hat{h}_{\mu\nu}~,
\end{equation}
\begin{equation}\label{1.2}
\hat{h}_{uu}=\hat{h}_{vv}= \frac{r_g}{2r}~,~ \hat{h}_{ij} = \delta_{ij} \frac{r_g}{r}~,~ \hat{h}_{ua}=\hat{h}_{vi}=0~,~
\end{equation}
\begin{equation}\label{1.3}
r^2 = (v-u)^2/4+ y_i^2~,~ r_g = 2 mG~,
\end{equation}
where $\eta_{\mu\nu}$ is the Minkowsky metric and $m$ is the mass of the particle.  If the massive particle is absent, $T_{(\mbox{\tiny{m}})}^{\mu\nu}=0$,  the gravitational field of the null particle is described by the metric of a plane-fronted shockwave (for a brief introduction to GSW, see \cite{Fursaev:2025wvh})
\begin{equation}\label{1.4}
g_{\mu\nu}=\eta_{\mu\nu}-Hl_\mu l_\nu~,
\end{equation}
where $l_\mu=-\delta^u_\mu$ are null normal to constant $u$ hypersurfaces, and
\begin{equation}\label{1.5}
H(v,u,y)=\chi(u) f(y)~,
\end{equation}
\begin{equation}\label{1.6}
\chi(u)=\delta(u)~,~ f(y)=8GE \ln (\rho/\rho_0)~,~\rho^2=(y_1+a)^2+y_2^2~~.
\end{equation}
(The dimensional parameter $\rho_0$ is fixed later.)
The hypersurface $u=0$ is the trajectory of the front of the shockwave. It is a null hypersurface since its normal vector $l$ is null, and it is the past event horizon of the null particle.
Function $f$ determines a 'profile' of the shockwave. Note that (\ref{1.4}) is an exact solution to the Einstein equations sourced by
\begin{equation}\label{1.6b}
T_{\mbox{\tiny{(n)}}}^{\mu\nu}=\chi(u)\sigma(y) l^\mu l^\nu~, ~\sigma(y)=E \delta^{(2)}(y)~,
\end{equation}
where $\sigma(y)$ is the surface energy density. When the null particle moves on a curved background geometry definition of the stress-energy should be modified. We discuss this topic in Appendix \ref{App1}.

As we have already noted, the collisions of null and massive particles are our primary interest. It is a simplest two-body system with a variety of applications. However, the approach which is presented below is applicable when the null particle is replaced with a null shell 
with an arbitrary profile $f=f(y)$. The stress-energy tensor of the null shell is given by (\ref{1.6b}) with $\sigma=f_{,ii}/\kappa$. The only condition we impose is that $f$ does not depend on advanced time $v$. In particular we can describe gravitational effects in the collisions when the null source is a null cosmic string (this problem has been already analyzed in \cite{Fursaev:2023oep}) or a null domain wall. In what follows it is convenient keep $f$ arbitrary until calculations of the concrete effects in Section \ref{sec:3}. 

The profile function encodes both the properties of the null source and of the created plane fronted GSW. The stress-energy tensor of the source 
does not depend on the traceless part of 2-tensor $f_{,ij}$. This traceless part yields 2 degrees of freedom associated to the polarizations
of the shockwave, see \cite{Barrabes:book}. There is a single nontrivial Newman-Penrose invariant
\begin{equation}\label{1.6bc}
\Psi_4=\frac 14 \delta(u)(f_{,11}-f_{,22}-2i f_{,12})~, 
\end{equation}
which solely depends on the polarization of the shockwave. 
The physical consequences of  (\ref{1.6bc}) can be inferred from Sachs equations when a null congruence crosses the shockwave.

There are restrictions when the linearized approximation we use is valid. We assume that distances from the massive and null particles to the region where effects are studied are much larger than gravitational radii $2mG$, $2EG$. It implies, in particular, that $a\gg mG$, $a\gg EG$.

\subsection{Gravitational memory}\label{sec:2.2}

In general, the shockwave generates perturbations $h_{\mu\nu}$ of the gravitational potential, so the total metric becomes
\begin{equation}\label{1.7}
g_{\mu\nu}=\eta_{\mu\nu}-Hl_\mu l_\nu+\hat{h}_{\mu\nu}+h_{\mu\nu}~.
\end{equation}
Examples when the shockwave from a null particle does not generate perturbations are well-known \cite{tHooft:1985NPB}, \cite{Sfetsos:1994xa} and depend 
on the background gravitational field. This is not our case.

Since the wave front  trajectory $u=0$ is the past event horizon of the null particle, the perturbations vanish in the region $u<0$. Therefore we look for a solution
\begin{equation}\label{1.8}
g_{\mu\nu}=\eta_{\mu\nu}+\hat{h}_{\mu\nu}+h_{\mu\nu}~,~u>0~,
\end{equation}
of the linearized Einstein equations (\ref{1.0}) in the future to the wave front.
The GSW does not contribute  explicitly to (\ref{1.8}) because $H=0$ at $u>0$.
However perturbations $h_{\mu\nu}$ do depend on the shockwave due to the gravitational memory.

The gravitational memory effect can be described for the `sandwich' type shockwave  (when 
$\chi(u)$ in (\ref{1.5})  is a smooth function which vanishes outside some narrow interval $0<u<\delta$; in the limiting case, $\delta \to 0$, $\chi(u)$ behaves as a delta-function). The effect of such a GSW on test particles and fields, just after the wave has gone can be described as coordinate transformations 
$\delta x^\mu=-\zeta^\mu(x)$ generated by the vector field
\begin{equation}\label{1.9}
\zeta^\mu(x)\simeq f\delta^\mu_v+\frac u2 \eta^{\mu a}f_{,a}~,
\end{equation}
where $x^a=(v,y^i)$. 
At $u=0$ transformation (\ref{1.9}) is reduced to the supertranslation 
\begin{equation}\label{1.10}
\delta x^a=-\hat{\zeta}^a=-f\delta^a_v~.
\end{equation}
Note that (\ref{1.9})  and results we report below hold for arbitrary profile functions $f=f(v,y)$.  For test freely moving particles with 4-velocities $U_-^\mu$ at $u<0$ the gravitational memory effect is the following variation 
of the velocities \cite{Fursaev:2025wvh}:
\begin{equation}\label{1.11}
\delta U^\mu\simeq {\cal L}_\zeta U_-^\mu~,~u=\delta~,
\end{equation}
where ${\cal L}_\zeta$ is the corresponding Lie derivative.
Analogous effects are true for non-gravitational fields. The gravitational memory of a free scalar field $\phi$ 
 is the variation
\begin{equation}\label{1.12}
\delta \phi \simeq {\cal L}_\zeta \phi_-~,~u=\delta~.
\end{equation}
Also variations of normal components of the Maxwell strength tensor look as coordinate transformations \cite{Fursaev:2025did}
\begin{equation}\label{1.13}
l^\mu  \delta F_{\mu\nu}\simeq l^\mu {\cal L}_\zeta (F_-)_{\mu\nu}
~,~u=\delta~.
\end{equation}
Here $\phi_-$ and $F_-$ are values of fields just before the action of the shockwave.

Variations (\ref{1.12}), (\ref{1.13}) allow one completely fix Cauchy data on the wave front (in the limit $\delta\to 0$) and find perturbations 
of scalar and Maxwell fields caused by the shockwave as solutions
to the following problems:
\begin{equation}\label{1.14a}
	\begin{cases}
		(-\Box+m^2) ~\phi=0~,~u>0
		\\
		\phi= {\cal L}_\zeta \phi_- ~,~u=0~,
	\end{cases}
	\qquad
	\begin{cases}
		\Box ~F_{\mu\nu}=0~,~u>0
		\\
		l^\mu  F_{\mu\nu}= l^\mu {\cal L}_\zeta (F_-)_{\mu\nu}
		~,~u=0~,
	\end{cases}
\end{equation}
see \cite{Fursaev:2024czx}, \cite{Fursaev:2025did}. Since the initial data are set on the null 
hypersurface, (\ref{1.14a}) belong to characteristic Cauchy problems.
The relevant Cauchy data for perturbations of velocities and fields are uniquely determined by (\ref{1.10}).

The case of the Cauchy problem for the Maxwell field is instructive for the analysis of the gravitational perturbations, since the free Maxwell equations result in the constraint $(\partial^\mu F_{\mu\nu})l^\nu=0$ at $u=0$ which does not involve derivatives over $u$.
One can check that supertranslations in the Cauchy data in (\ref{1.14a}) are consistent with the constraint. It also implies that the Cauchy data for the Maxwell field (if one takes into account the Bianchi identities) are fixed by the two components $F_{vi}$. 

As we see, an analogous consistency property between the constraints and supertranslations holds for metric perturbations caused by the GSW.

\subsection{Soldering geometries across the shockwave front}
\label{sec:2.3}

The gravitational memory effect described above can be reinterpreted in a more geometric  way by using the approach by Penrose  \cite{Penrose:1972xrn} and the null-shell formalism \cite{Barrabes:1991ng}, \cite{Poisson:2002nv}. 
The discussion of this Section concerns shockwaves with arbitrary profiles $f$.
Consider a manifold $\cal M$ and a null hypersurface  ${\cal N}$  which partitions $\cal M$  into two parts, ${\cal M}^\pm$ with corresponding metrics $g^{\pm}_{\mu\nu}$. Our convention is that  ${\cal M}^+$ is in the future of $\cal N$ and 
${\cal M}^-$ is in the past of $\cal N$.
We choose coordinate charts $x^\mu_\pm$ on ${\cal M}^\pm$ and coordinates $x^a$ on $\cal N$.

Our interest is in the case when components of the Riemann tensor of $\cal M$ have delta-function-like singularities on $\cal N$.
In the null-shell approach this happens when normal derivatives of $g^{\pm}_{\mu\nu}$ do not coincide on $\cal N$. The jump of derivatives determines the surface energy, pressure and  current of the shell defined by components of the so called transverse curvature
\cite{Barrabes:1991ng}, \cite{Poisson:2002nv}. 

Suppose now that $\cal M$ is locally $R^{1,3}$, and $\cal N$ is the trajectory of the front $u=0$.  Let $x^\mu$ be {\it Minkowsky coordinates}, and $U^\mu$ be components of 4-velocities in coordinates $x^\mu$ in the absence of the shockwave.
If we require that coordinate charts and components of the 4-velocities are continuous across the wave front,
the gravitational memory (\ref{1.11}) implies that
\begin{equation} \label{2.1a}
x^{\mu}_+=x^\mu~,~u>0~,
\end{equation}
\begin{equation} \label{2.1b}
x^{\mu}_-=x^\mu-\zeta^\mu(x)~,~u<0~,
\end{equation}
\begin{equation} \label{2.1}
x^{\mu}_+=x^{\mu}_-~,~u=0~,
\end{equation}
\begin{equation} \label{2.2}
U^{\mu}_+=U^{\mu}_-\simeq U^{\mu} + {\cal L}_\zeta U^\mu~,~u=0~,
\end{equation}
where $\zeta$ is defined in (\ref{1.9}), and coordinates $x^\mu$ are assumed to be discontinuous. Thus, the gravitational memory relations are ensured if one uses on ${\cal M}^-$ the "curved" coordinates $x_-$. The Penrose approach is to glue two halfs of the Minkowsky spacetime, ${\cal M}^\pm$, but not along flat coordinate charts. This also implies that \cite{Blau:2015nee}
\begin{equation} \label{2.3}
g^{+}_{\mu\nu}(x_+)=\eta_{\mu\nu}~,~g^{-}_{\mu\nu}(x_-) \simeq \eta_{\mu\nu}+{\cal L}_\zeta \eta_{\mu\nu}(x_-)~,~u<0~.   
\end{equation}
The vector field (\ref{1.9}) has the property
\begin{equation} \label{2.3a}
{\cal L}_\zeta \eta_{\mu\nu}=\zeta_{\mu,\nu}+\zeta_{\nu,\mu}=0~,~u=0~,   
\end{equation}
which guarantees that the metric components are continuous, $g^{+}_{\mu\nu}=g^{-}_{\mu\nu}$, across $\cal N$. 

The normal derivatives of the metric are not continuous.
One can calculate the surface energy of the null-shell and check  it coincides with the energy of the null source of the shockwave, see (\ref{2.18}) and \cite{Fursaev:2023oep} for further details.

We extend the above analysis to the case of shockwaves on non-flat manifolds in Section \ref{sec:2.4} by working in the linearized approximation,
\begin{equation}\label{3.2}
g^{\pm}_{\mu\nu}=\eta_{\mu\nu}+h^{\pm}_{\mu\nu}~,
\end{equation} 
where $h^{\pm}_{\mu\nu}$ are small perturbations. The present Section provides the necessary constructions.  From now on we start to use the single notation $x^\mu=(u,v,y^i)$ for the continuous coordinates $x^\mu_\pm$.  The notation holds across $\cal N$. The Latin indexes $a,b$ will numerate the three coordinates $v,y^i$.

It is convenient to choose coordinates where equation of the  shockwave front trajectory $\cal N$ is still $u=0$.  In the chosen coordinates 
the metric tensor of $u=0$ coincides with $g_{ab}$ and we
require that these components are continuous across $\cal N$ ,
\begin{equation} \label{2.6}
[g_{ab}]=[h_{ab}]=0~,~u=0~.
\end{equation} 
The notation $[A](x)\equiv A^+(x)-A^-(x)$, where $A_{\pm}=\lim_{u\to 0\pm}A$, will be used to measure discontinuity of a quantity $A$. Additionally, we demand that 
\begin{equation} \label{2.5}
g^{\pm}_{va}=h^{\pm}_{va}=0~,~u=0~.
\end{equation} 
These conditions, in the given approximation,  ensure that the normal vector $\partial_\mu u$ is null, 
and the curves lying on $u=0$ at fixed $y^i$  are null geodesics generated by the null vector $\partial_v$, see more details in Appendix \ref{App1}.   

Since the trajectory of the null particle belongs to $\cal N$, the future null cones with the origin on the trajectory are tangent to $\cal N$. Thus, $\cal N$ is the past event horizon of the null particle. It is in this sense, we identify $\cal N$ with the Cauchy hypersurface for the perturbations caused by the shockwave. In general, $\cal N$ may be a part of a true null Cauchy hypersurface of the spacetime.

Condition (\ref{2.6}) does not guarantee that the geometry is smooth. To relate, in the linearized theory, the surface discontinuities of the geometry to properties of the null matter on $\cal N$  we use definitions which follow from the variational principle
of the gravitational action.
The total action of the problem we study is
\begin{equation} \label{2.7}
I = I_{\mathrm{EH}} +  I_{\mathrm{shell}}+I_{\mbox{\tiny{(m)}}}+I_{\mbox{\tiny{(n)}}}~.
\end{equation}
Here $I_{\mathrm{EH}}$ is the standard Einstein-Hilbert action defined on ${\cal M}^+\cup {\cal M}^-$,
\begin{equation}
I_{\mathrm{EH}} =\frac{1}{\kappa} \int d^4x \sqrt{-g} R~.
\end{equation}
$I_{\mbox{\tiny{(m)}}}$ is the action of matter with non-zero mass (a massive particle, for example), $I_{\mbox{\tiny{(n)}}}$ is the action of a null matter which creates the shockwave and is defined on $u=0$. 

An additional term $I_{\mathrm{shell}}$ is needed to ensure that variations of the gravity action on the Einstein equations 
depend only on variations of the metric on $\cal N$,
\begin{equation} \label{2.8}
\delta\left[I_{\mathrm{EH}} +  I_{\mathrm{shell}}\right]=-\frac{1}{2\kappa} \int_{u=0}d^3x~ C^{ab}(h^+,h^-)\delta h_{ab}~.
\end{equation}
In the linearized theory one finds (see Appendix  \ref{App1} for our conventions and other details) that 
\begin{multline}\label{2.9}
	I_{\mathrm{shell}} = \frac{1}{\kappa} \int_{u=0}d^3x~ \bigl( h_{ii} [\partial_u h_{vv}] +4 [h_{uv}\partial_u h_{vv}]- 4 h_{vi}~  [\partial_u h_{vi}]+ 2h_{vv} ~[\partial_u h_{ii}]
	\\
	 -  2[h_{uu}]~ \partial_v h_{vv}  -2[h_{uv}] (2 \partial_i h_{vi} - \partial_v h_{ii})  +2[h_{ui}]~ \partial_i h_{vv} \bigr)~, 
\end{multline}
where we used condition (\ref{2.6}).  By taking this into account one obtains $C^{ab}$ in on-shell variation (\ref{2.8})
\begin{equation}\label{2.10}
	\begin{split}
		C^{vv} (h^+,h^-) &= 2(2 [\partial_i h_{ui}]-[\partial_u h_{ii}]) ~, 
		\\
		C^{vi}(h^+,h^-) & = 2([\partial_u h_{vi}]-[\partial_i h_{uv}] -[\partial_v h_{ui}])~, 
		\\
	    C^{ij}(h^+,h^-) &= 2\delta_{ij} ( 2 [\partial_v h_{uv} ]- [\partial_u h_{vv}] )~.
	\end{split}
\end{equation}
Note that the 2-index structure $C^{ab}$ is not a 3-tensor on $\cal N$ but rather a tensor density. Definition (\ref{2.8}) differs from the standard definition of a tensor by variation of the action functional since the metric of $\cal N$ is degenerate. In what follows we call $C^{ab}$ the surface curvature density.
Analogously, we call
\begin{equation}\label{2.12}
t^{ab} = \frac{\delta I_{\mbox{\tiny{(n)}}}}{\delta h_{ab}} ~,
\end{equation}
the surface stress tensor density of the null source which creates the shockwave.  By this definition, it is related to components of 4-dimensional stress-energy tensor (\ref{1.6b}) as
\begin{equation}\label{2.12b}
T_{\mbox{\tiny{(n)}}}^{ab}(x) =\delta(u)\frac{2}{\sqrt{-g}}~t^{ab}(x^a)~.
\end{equation}
By virtue of (\ref{2.5}),  $g=-(g_{uv})^2\det{g_{ij}}$ on $\cal N$. As we see, $[g]=0$.
We require that variations of total gravity action 
(\ref{2.7})  
on the Einstein equations under conditions (\ref{2.6}) vanish. This yields the important connection
\begin{equation}\label{2.13}
2C^{ab}=\kappa~ t^{ab}~
\end{equation}
between the surface curvature and surface energy on the null-shell.

The surface curvature is related to the singular part of components of the Einstein tensor on the shell as 
\begin{equation}\label{2.16}
G^{ab}\simeq \delta(u) C^{ab}~, 
\end{equation}
see Appendix  \ref{App1} for details.  From the Einstein equations one finds at $u\to 0$
\begin{equation}\label{2.15}
2\partial _a G^{ab}= -\kappa~l_\mu T_{\mbox{\tiny{(m)}}}^{\mu b}~.
\end{equation}
Since the matter flow $l_\mu T_{\mbox{\tiny{(m)}}}$ through $\cal N$ is continuous and does not have singularities like $\delta(u)$
one concludes that
\begin{equation}\label{2.17}
2\partial_a C^{ab}=-\kappa~\lim_{\varepsilon \to 0}\int_{-\varepsilon}^{\varepsilon}l_\mu T_{\mbox{\tiny{(m)}}}^{\mu b}~du=0~.
\end{equation}
Equations (\ref{2.13}), (\ref{2.17})  are consistent with the surface conservation law $\partial_a t^{ab}=0$.

In the null-shell approach one also defines the transverse curvature \cite{Barrabes:1991ng}, \cite{Poisson:2002nv}
\begin{equation}\label{2.11.1}
\mathbb{C}_{ab}= e_a^\mu e_b^\nu \nabla_\nu n_\mu~,
\end{equation}
where $e_a=(l,e_i)$, $e_i$ are space-like and tangent to $\cal N$, vector $n$ is null $(l\cdot n)=-1$, $(n\cdot e_i)=0$. 
In the linearized theory the surface curvature and the transverse curvatures are related as
\begin{equation}\label{2.11}
C^{vv}= -4 [\mathbb{C}_{ii}]~,~ 
C^{vi} = 2[\mathbb{C}_{vi}]~,~
C^{ij} = -\delta^{ij} [\mathbb{C}_{vv}] ~,
\end{equation}
if one takes $n=\partial_u$, $e_i=\partial_i$.
We provide (\ref{2.11.1}), (\ref{2.11}) just to make a link with Israel-Barrabes approach. We do not use $\mathbb{C}_{ab}$ in what follows,
except Section \ref{sec:3.4}.

For the plane-fronted shockwave in the Minkowsky spacetime the surface curvature $C^{ab}$ is non-vanishing since
$h^{-}_{\mu\nu}=\zeta_{\mu,\nu}+\zeta_{\nu,\mu}$, see (\ref{2.3}). The straightforward computation yields
\begin{equation}\label{2.18}
C^{vv}(0,{\cal L}_\zeta \eta)=2f_{,ii}~,~ 
C^{vi}(0,{\cal L}_\zeta \eta)= - 2f_{,vi}~,~
C^{ij}(0,{\cal L}_\zeta \eta)= 2\delta^{ij} f_{,vv}~.
\end{equation}
There are infinitely many ways to glue continuously two halves of the Minkowsky spacetime along the null hypersurface $u=0$.
All these ways, however, are physically different, as a consequence of (\ref{2.18}).

\subsection{Transverse curvature, constraints and Cauchy data}
\label{sec:2.4}

We return now to problem of the gravitational field of the double system consisting of null and massive particles.
To use results of the previous Section we need to find deformation of the null hypersurface and trajectory of the null particle under the gravitational field of the massive particle.  In the linearized theory we change coordinates $u$, $y^i$ and components  
$\hat{h}_{\mu\nu}$,  see (\ref{1.2}), to ensure condition (\ref{2.5}) in the transformed coordinates, see Appendix \ref{App1}.

Our convention in (\ref{1.2}), (\ref{1.3})  is that the massive particle crosses $\cal N$ at the point $x^a=0$.
For technical reasons it is convenient to assume that 
\begin{equation}\label{3.0}
f(0)=f_{,i}(0)=0~.
\end{equation}
Conditions (\ref{3.0}) can be always achieved by a Lorentz transformation which looks as a shift $f(x)$ to $f(x)-f(0)-f_{,i}(0)y^i$.
This shift does not affect the surface curvature  (\ref{2.18}), see \cite{Fursaev:2024czx} for details.  Equations (\ref{3.0}) also 
fix dimensional parameter $\rho_0$ in (\ref{1.6}).

In the absence of the null particle the linearized Einstein equations are
\begin{equation}\label{3.1}
2G^{\mu\nu}(h)= \kappa~ T_{\mbox{\tiny{(m)}}}^{\mu\nu}~,~g_{\mu\nu}=\eta_{\mu\nu}+\hat{h}_{\mu\nu}~.
\end{equation}
As has been explained we consider a null hypersurface $\cal N$ which partitions $\cal M$  into ${\cal M}^+$ and ${\cal M}^-$ with metrics (\ref{3.2}). The trajectory of a null particle is a null geodesic which belongs to $\cal N$. The future directed null cones with tips on the particle's trajectory are tangent to $\cal N$ and lie in ${\cal M}^+$. Therefore $\cal N$ is the past event horizon of the particle.  The perturbations $h_{\mu\nu}$ caused by the null particle may exist only in ${\cal M}^+$.

In the chosen coordinates we identify the trajectory of the null particle by the same equations as in the flat case, $u=y_2=0$, $y_1=a$ .
In accord with our arguments based on the gravitational memory effect the metric on ${\cal M}^-$ is the coordinate transform
\begin{equation} \label{3.3}
g^{-}_{\mu\nu}=g_{\mu\nu}+{\cal L}_\zeta g_{\mu\nu}~,~u<0~,  
\end{equation}
generated by supertranslation (\ref{1.9}). Therefore
\begin{equation} \label{3.4}
h^{-}_{\mu\nu}=\zeta_{\mu,\nu}+\zeta_{\mu,\nu}+\hat{h}_{\mu\nu}+{\cal L}_\zeta \hat{h}_{\mu\nu}~,~u<0~,
\end{equation}
where $\zeta_\mu=\eta_{\mu\nu}\zeta^\nu$.
The problem in ${\cal M}^+$ is the following:
\begin{equation}\label{3.5a}
2G^{\mu\nu}(h^+)= \kappa~ T_{\mbox{\tiny{(m)}}}^{\mu\nu}~,~u>0~,
\end{equation}
\begin{equation}\label{3.5b}
h^+_{\mu\nu}=\eta_{\mu\nu}+\hat{h}_{\mu\nu}+h_{\mu\nu}~,~u>0~.
\end{equation}
Note that $h^+_{\mu\nu}$ includes the perturbations $h_{\mu\nu}$ of the gravitational field which may be generated by the null particle.
The need in perturbations is dictated by
the continuity conditions on $\cal N$
\begin{equation}\label{3.7}
h^+_{ab}=h^-_{ab}~,~u=0~,
\end{equation}
which must be imposed. Equation (\ref{3.7}) and the fact that $\zeta_{\mu,\nu}+\zeta_{\mu,\nu}$ vanishes require that $h_{ab}={\cal L}_\zeta \hat{h}_{ab}$ on $\cal N$. Thus, perturbations are non-trivial once ${\cal L}_\zeta \hat{h}_{ab}\neq 0$, and supertranslations symmetries are violated on $\cal N$.  We return to this property in Section \ref{sec:2.5}.

The equations (\ref{3.5a}) must guarantee that the surface curvature density is correctly related to the surface stress tensor 
of the null particle,
\begin{equation}\label{3.6}
2C^{ab}(h^+,h^-)=\kappa~ t^{ab}~.
\end{equation}
Definition of $t^{ab}$ in an arbitrary background field is discussed in Appendix \ref{App1}. The only non-vanishing
component of $t^{ab}$ is $t^{vv}$ which can be written in the linear order as 
\begin{equation}\label{3.12}
t^{vv}=\tilde{f}_{,ii}/\kappa~,~\tilde{f}(y,v)=(1+4\hat{h}_{uv}(v,y_0))f(y)~,
\end{equation}
where $f$ is defined in (\ref{1.6}). This means that the gravitational field of the massive particle yields a $v$-dependent correction to the profile function $f$ of the order $O(\hat{h} f)$, and it goes beyond the considered approximation. Therefore, we use  for $t^{ab}$ in (\ref{3.6}) the same expression as in the Minkowsky background. 

We are interested in the perturbation $h_{\mu\nu}$ caused by the GSW. By taking into account (\ref{3.1}), (\ref{3.4}) one can formulate the Cauchy problem
\begin{numcases}{}
  G^{\mu\nu}(h)=0~,~u>0~, \label{3.8}
  \\
  h_{ab}={\cal L}_\zeta \hat{h}_{ab}~,~u=0~, \label{3.9}
  \\
  C^{ab}(h,{\cal L}_\zeta \hat{h})=0~. \label{3.10}
\end{numcases}
Condition (\ref{3.10}) follows from (\ref{3.6}),  linearity of $C^{ab}$, which implies that
\begin{equation}\label{3.11}
C^{ab}(h^+,h^-)=C^{ab}(0,{\cal L}_\zeta \eta)+C^{ab}(h,{\cal L}_\zeta \hat{h})~,
\end{equation}
and from the fact that $C^{ab}(0,{\cal L}_\zeta \eta)=8\pi G~ t^{ab}$, see  (\ref{2.18}). If $\hat{h}_{vv}=0$ on $u=0$, then $h_{vv}=0$ on $u=0$. Einstein equations require (\ref{2.17}). In combination with (\ref{3.11}) they result in the conditions:
\begin{equation}\label{3.17}
\partial_a C^{ab}(h,{\cal L}_\zeta \hat{h})=0~.
\end{equation}
The consequence of (\ref{3.17}) is that (\ref{3.10}) hold if 
\begin{equation}\label{3.18}
C^{ij}(h,{\cal L}_\zeta \hat{h})=0~~~\mbox{or}~~~2 [\partial_v h_{uv} ]- [\partial_u h_{vv}]=0~,
\end{equation}
where $[h_{\mu\nu}]\equiv h_{\mu\nu}-{\cal L}_\zeta \hat{h}_{\mu\nu}$.
Indeed, it follows from (\ref{3.18}) that $\partial_v C^{vi}(h,{\cal L}_\zeta \hat{h})=0$ and then  $ C^{vi}(h,{\cal L}_\zeta \hat{h})=0$
if perturbations $h_{\mu\nu}$ vanish at past null infinity, $v\to -\infty$. Analogously for $C^{vv}(h,{\cal L}_\zeta \hat{h})$.

It should be noted that equation $G_{vv}(h)=0$ does not contain derivatives over $u$. Therefore, this equation, taken on $\cal N$, is the constraint on data (\ref{3.9}).  Given condition (\ref{3.9}) this constraint is satisfied,
\begin{equation}\label{3.12.1}
\left. G_{vv}(h)\right |_{u=0}=G_{vv}({\cal L}_\zeta \hat{h})={\cal L}_\zeta G_{vv}(\hat{h})+\frac 12 f_{,ii}\partial_v \hat{h}_{vv}
=\frac{\kappa}{2} ~{\cal L}_\zeta (T_{\mbox{\tiny{(m)}}})_{vv}(\hat{h})=0~,
\end{equation}
since ${\cal L}_\zeta (T_{\mbox{\tiny{(m)}}})(\hat{h})=f(0)\partial_v (T_{\mbox{\tiny{(m)}}})_{vv}(\hat{h})=0$, as a result of (\ref{3.0}).

The coordinate transformations $h'_{\mu\nu}=h_{\mu\nu}-\xi_{\mu,\nu}-\xi_{\nu,\mu}$, where $\xi_\mu=0$ on $u=0$, leave 
initial data (\ref{3.9}), (\ref{3.10}) invariant. Conditions $\xi_a=0$ are required by (\ref{3.9}), while $\xi_u=0$ since supertranslations are already fixed by   (\ref{3.16}).   
We now demonstrate that data (\ref{3.9}), (\ref{3.10})
allow one to find a solution $h_{\mu\nu}$ up to the above coordinate transformations. 

In what follows it is convenient to use the de Donder gauge
\begin{equation}\label{3.13}
\partial_\mu \bar{h}^\mu_\nu=0~,~\bar{h}_{\mu\nu}=h_{\mu\nu}-\frac 12 \eta_{\mu\nu} h~,
\end{equation}
where $h=\eta^{\mu\nu} h_{\mu\nu}$ and indices are risen and lowered with the flat metric. If $\partial_\mu \bar{h}^\mu_\nu=\psi_\nu\neq 0$ one can use the coordinate freedom to impose (\ref{3.13}). The vector field $\xi$ is uniquely determined by the Cauchy problem
\begin{equation}\label{3.14}
\Box \xi_\mu=\psi_\mu~,~u>0~;~ \xi_\mu=0~,~u=0~.
\end{equation}
In the de Donder gauge  equations (\ref{3.8}) become $\Box h_{\mu\nu}=0$ and imply that $h_{va}\equiv 0$.

Initial data (\ref{3.9}) allow one to find solutions for components $h_{ab}$.  
One needs initial data for  the rest 4 components $h_{u\mu}$. To this aim we have 4 gauge conditions (\ref{3.13}) but condition 
\begin{equation}\label{3.16}
\partial_\mu \bar{h}^\mu_v=-2h_{vv,u}+h_{vi,i}-\frac 12 h_{ii,v}=0~
\end{equation}
does not depend on  $h_{u\mu}$. This condition is related to the constraint, $\partial_v\partial_\mu \bar{h}^\mu_v=G_{vv}$.
(Since $h_{va}\equiv 0$, the byproduct of  (\ref{3.16}) is that $h_{ii}\equiv 0$.) 

Thus, we need additional inputs to fix data,  for example, for $h_{uv}$. Once solution for $h_{uv}$ is known, the data for $h_{uu}$ 
and $h_{ui}$ can be inferred from the gauge conditions.  This additional information does exist and it is related to the demand that surface curvature $C^{ab}$ satisfies (\ref{3.6}).  As has been mentioned above, equations (\ref{3.10}) are reduced to the single condition 
(\ref{3.18}) which in the chosen gauge takes the form:
\begin{equation}\label{3.19}
\partial_v [h_{uv}]=0~,~u=0~.
\end{equation}
To get it one should use (\ref{3.16}) and the fact that $h_{vv}\equiv 0$. If we require that perturbations vanish at $v\to -\infty$ then (\ref{3.19}) just implies the additional initial condition
\begin{equation}\label{3.20}
h_{uv}={\cal L}_\zeta \hat{h}_{uv}~,~u=0~.
\end{equation}
To summarize results of this Section, it has been demonstrated that the characteristic Cauchy problem (\ref{3.8})-(\ref{3.10}) for perturbations
$h_{\mu\nu}$ of the gravitational field $\hat{h}_{\mu\nu}$ of a massive particle at rest, which are caused by the gravitational shockwave of a null particle, is equivalent to the following Cauchy problem: 
\begin{numcases}{}
	\Box h_{\mu\nu}=0~,~u>0~, \label{3.15}
	\\
	h_{uv}={\cal L}_\zeta \hat{h}_{uv}~,~h_{ab}={\cal L}_\zeta \hat{h}_{ab}~,~u=0~, \label{3.21}
	\\
	\partial_\mu \bar{h}^{\mu a}=0~,~u=0~. \label{3.22}
\end{numcases}
Conditions (\ref{3.22}) fix initial data for 3 components $h_{uu}$, $h_{ui}$. Below we show how to find a unique solution for the perturbations in the de Donder gauge.

\subsection{Gravitational waves from shockwaves and breakdown of supertranslations }
\label{sec:2.5}

As has been pointed out earlier, gravitational shockwaves produced by null particles in background gravitational fields have been studied since the seminal work \cite{tHooft:1985NPB} and used in numerous applications.  Some generalizations have been suggested latter in \cite{Sfetsos:1994xa}. Here we explain why shockwaves in these papers do not produce perturbations of background fields and produce no gravitational waves, see Section
\ref{sec:3}.

Equation (\ref{3.21}) determines a necessary condition for the shockwave to generate perturbations of the background gravitational fields.
Perturbations exist if the Cauchy data for $h_{uv}$ and $h_{ab}$ are non-trivial. As is easy to see, this happens when
\begin{equation}\label{3.21b}
{\cal L}_\zeta g_{uv}\neq 0~,~{\cal L}_\zeta g_{ab}\neq 0~,~u=0~,
\end{equation}
at least in the linearized approximation.  Thus, {\it the necessary condition for the perturbations is the violation of the supertranslation symmetries on} $\cal N$.

One can check that in the chosen coordinates, where $g_{av}=0$, Eq. (\ref{3.21b}) are equivalent to conditions
\begin{equation}\label{3.21c}
\partial_v g_{uv}\neq 0~,~\partial_v g_{ij}\neq 0~,~u=0~,
\end{equation}
which is the case of this paper. Early works \cite{tHooft:1985NPB}, \cite{Sfetsos:1994xa} had studied null particles in background geometries with conditions  $\partial_v g_{uv}=\partial_v g_{ij}=0$ on $\cal N$. The example is null particles which move along  Killing horizons.
This restriction explains why shockwaves from such particles did not generate gravitational perturbations.

A more general class of geometries and supertranslations which allows soldering ${\cal M}^+$ with  ${\cal M}^-$ without perturbations of 
${\cal M}^+$ is discussed in  \cite{Blau:2015nee}.

\section{Gravitational effects in the fixed-target collisions}
\label{sec:3}
\setcounter{equation}0

\subsection{The method to solve the Cauchy problem}\label{sec:3.1}

The aim of the rest paper is to solve the Cauchy problem (\ref{3.15})--(\ref{3.22}) for gravitational perturbations
in the background geometry with $\hat{h}_{\mu\nu}$ defined in \eqref{1.2}. We study specific properties of the solutions
and related physical effects. We follow conventions adopted in the previous Section. 
Coordinate transformations $\hat{h}_{\mu\nu}'=\hat{h}_{\mu\nu}+\chi_{\mu,\nu}+\chi_{\nu,\mu}$, which preserve the de Donder gauge and bring components to the form which respects conditions (\ref{2.5}) are discussed in Appendix \ref{App1}. Components of vector $\chi_\mu$ are defined by problem (\ref{a1.13}), (\ref{a1.14}).

We proceed by following the method elaborated in \cite{Fursaev:2024czx}.  Any profile function $f(y)$ can be written as the superposition of “elementary
profiles” $f_p(y)$,
\begin{equation} \label{4.3}
	f(y) = \frac{1}{(2\pi)^2} \int d^2p~  \tilde{f}(p) f_p(y)~,
\end{equation}
\begin{equation} \label{4.4}
	f_p(y) = e^{i p_j y^j}~,
\end{equation}
where $d^2p=dp_1 dp_2$.  In particular, for the shockwave from the null particle, see \eqref{1.6},
\begin{equation}\label{6.2}
	\tilde{f}(p) = -2\pi \omega \frac{e^{i p_1 a}}{p_i^2}~,~ \omega = 8GE~. 
\end{equation}
In the linearized theory, if $h_{\mu\nu}(x,p)$ are gravitational perturbations for $f_p(y)$,   one can write the general solution as
\begin{equation} \label{4.1}
h_{\mu\nu}(x) = \frac{1}{(2\pi)^2} \int d^2p~ \tilde{f}(p)~h_{\mu\nu}(x,p)~,
\end{equation}
in accord with (\ref{4.3}). It is more convenient to find perturbations $h_{\mu\nu}(x,p)$, and then use (\ref{4.1}) to get a solution
for the null particle or any other ultrarelativistic object.  

"Elementary perturbations" $h_{\mu\nu}(x,p)$ are solutions to problem (\ref{3.15})--(\ref{3.22}). To find their Cauchy data
we chose frame of reference in ${\cal M}^+$  where the massive particle is at rest in the center of coordinates. This requires to change the profile function $f(x)$ to $f(x)-f(0)-f_{,i}(0)y^i$, see discussion after (\ref{3.0}). Equivalently, the elementary profile $f_p(y)$ can be replaced with $\bar{f}_p(y) = f_p(y) - 1-i p_j y^j$. After these preliminaries the initial data \eqref{3.21} for non-vanishing components take the form
\begin{equation} \label{4.5}
h_{uv}(x,p) =   \frac{1}{2}\bar{f}_p \partial_v \hat{h}_{vv} 
		~,~u=0~,
\end{equation}
\begin{equation} \label{4.5.1}
h_{ij} (x,p)=\bar{f}_p \partial_v (2 \delta_{ij}  \hat{h}_{vv} +  \partial_i \chi_j+\partial_j \chi_i)~,~u=0~.
\end{equation}
 The rest non-trivial components, $h_{uu}$, $h_{ui}$ will be found from the gauge conditions.

A general solution to the characteristic Cauchy problem can be written in an integral form derived in  \cite{Fursaev:2024czx} for scalar perturbations caused by the shockwave in the field of a scalar point-like source (see Appendix A of \cite{Fursaev:2024czx}). Since the Cauchy problem for tensor components has the form of that for scalar fields, we can use these results to write the following solution
\begin{equation}  \label{4.7}
h_{\mu\nu}(x,p) =     \int_{S^2} d \Omega'~ \Theta(x \cdot \mathbf{l})~ \beta_{\mu\nu}(p,n) ~   e^{i k_+  x\cdot \mathbf{l}}~,
\end{equation}
Integration in (\ref{4.7}) goes over a unit sphere, $n_v^2+n_1^2+n_2^2=1$, with $n_v =\sin\theta' \cos\phi'$, $n_1 = \cos\theta'$, $n_2 = \sin\theta'\sin\phi'$ and $d\Omega'=\sin\theta' d\theta' d\phi'$. Other notations  include
\begin{equation}\label{4.11}
k_\pm \equiv k_\pm (p,n) =p_i n_i \pm i \sqrt{p_i^2-(p_i n_i)^2}~,
\end{equation}
and a vector $\mathbf{l}=\mathbf{l}(n)$ with the components
\begin{equation} \label{4.6}
	l_u= n_i^2/2 n_v~,~l_v = n_v/2~,~l_i = n_i~.
\end{equation}
Note that $\mathbf{l}$ is null, $\mathbf{l}^2=0$. This property guarantees that $\Box h_{\mu\nu}(x,p)=0$. The convergence of the integral 
in (\ref{4.7}) for all points $x$ and  momenta $p$ is ensured by the exponential factor.

The Cauchy data for $h_{uv}$ and $h_{ij}$ allow one to find
\begin{equation} \label{4.8}
	\beta_{uv} =     -\frac{ S_{vv}  }{k_+ l_v}~,~
	\beta_{ij} =  -\frac{2 ( S_{ij} +2\delta_{ij} S_{vv}) }{k_+ l_v}~,
\end{equation}
\begin{equation} \label{4.10}
	S_{ab}\equiv S_{ab}(p,n)= \frac{r_g}{4\pi}  \frac{ k_+^2 (k_+ l_a-p_a)(k_+ l_b-p_b)   }{k_+-k_-}~,~ p_v=0~,
\end{equation}
where we took into account (\ref{4.5}), (\ref{4.5.1}).  Note that form of $S_{ab}$ in (\ref{4.10}) is dictated by components $\hat{h}_{\mu\nu}$ 
which have a form of the Newton potential, see more details in \cite{Fursaev:2024czx}.

The gauge conditions  \eqref{3.22} are reduced to
\begin{equation} \label{4.9}
	l^\mu\left( \beta_{\mu\nu} - \frac{1}{2}\eta_{\mu\nu} \beta \right) = 0~,~
	\beta = \eta^{\mu\nu} \beta_{\mu\nu}=-4 \beta_{uv}~,
\end{equation}
and allow one to find
\begin{equation}
	\beta_{ui} =\frac{l_j \beta_{ij} +2 l_i \beta_{uv}}{2l_v} 
	~,~
	\beta_{uu} = \frac{l_i \beta_{ui} }{2l_v} 
	~. 
\end{equation}
Thus, all quantities $\beta_{\mu\nu}(p,n)$ in (\ref{4.7}) are fixed. The solution $h_{\mu\nu}(x)$ for the given profile follows when 
 (\ref{4.7}) is substituted in (\ref{4.1}). The final integral formula for perturbations is complicated. However, as will be shown in next Sections, there are two spacetime regions where an analytical treatment of physical effects is possible.

\subsection{The gravitational radiation in collision of null and massive particles}\label{sec:3.2}

One can suggest that perturbations $h_{\mu\nu}(x)$ behave as gravitational waves far from the massive particle. If we introduce
the radial coordinate $r=\sqrt{x^2+y_i^2}$, centered at the particle's position, and the retarded time $U=t-r$, then large distances correspond to large $r$ at finite $U$. Thus we study large $r$ (or late-time) asymptotics of $h_{\mu\nu}(x)$.
To this aim it is convenient to go to coordinates
\begin{equation}\label{4.11.1}
ds^2 = -dU^2-2dU dr+r^2 d\Omega^2~,
\end{equation} 
where $d\Omega^2 = \gamma_{AB} dx^A dx^B = \sin^2 \theta d\phi^2+d\theta^2$.
In coordinates (\ref{4.11.1}) the asymptotic $r\to \infty$ describes the future null infinity $\mathcal{J}^+$. 
Notation $x^A$ stands for spherical coordinates $\theta,\phi$, metric $\gamma_{AB}$ is the metric on the unit sphere. 
We will also use a unit vector $\vec{e}$ to define the direction from the massive particle to a distant observer with coordinates $x,y^i$
 \begin{equation} \label{4.13}
 \vec{e}(\Omega)= \frac{\vec{x}(\Omega)}{r}~.
 \end{equation}
Calculations show, see details in Appendix \ref{App2}, that the solution  behaves at $r \to \infty$ as 
\begin{equation}\label{4.15}
	h_{\mu\nu}(x,p) =\frac{H_{\mu\nu}(U,\Omega,p )}{r}   + O(r^{-2})~,
\end{equation}
\begin{equation}\label{4.17a}
H_{\mu\nu}^+ (U,\Omega,p )\equiv H_{\mu\nu}(U,\Omega,p )\Bigl|_{U>0}=
	\frac{i\pi  \tilde{\beta}_{\mu\nu} }{\tilde{k}_+}  
	~ \exp\left( \frac{i \tilde{k}_+  U}{ 2 \tilde{n}_v}      \right)
~,
\end{equation}
\begin{equation}\label{4.17b}
H_{\mu\nu}^- (U,\Omega,p )\equiv H_{\mu\nu}(U,\Omega,p )\Bigl|_{U<0}=-
	\frac{i \pi  \tilde{\beta}^*_{\mu\nu} }{\tilde{k}_-}  
	~ \exp\left( \frac{i \tilde{k}_-  U}{ 2 \tilde{n}_v} \right)+2\pi i~ \mathrm{Re} \left(\frac{\tilde{\beta}_{\mu\nu} }{\tilde{k}_+}\right)~,
\end{equation} 
\begin{equation}\label{4.19}
\tilde n_v(\Omega)=\sqrt{\frac{1-e_x}{2}}~,~\tilde n_i(\Omega)=-\frac{e_i}{\sqrt{2(1 -e_x)}}~,
\end{equation}
\begin{equation}\label{4.19b}
\tilde{\beta}_{\mu\nu}(p,\Omega)\equiv\beta_{\mu\nu}(p,\tilde{n})~,~\tilde{k}_\pm (p,\Omega) \equiv k_\pm (p,\tilde{n})~.
\end{equation}
Here the components are given in the Minkowsky coordinates.  In addition to dynamic terms the asymptotics (\ref{4.15}) also include
static terms of the same order $O(1/r)$. These terms, which are given in  Appendix \ref{App2}, see (\ref{4.18}), are not important for the subsequent analysis. The stationary-point method which we used to get (\ref{4.17a}), (\ref{4.17b}) requires that $e_x < 1$. However, results
(\ref{4.17a}), (\ref{4.17b}) are well-defined at $e_x=1$.  

The amplitudes $H_{\mu\nu}$ are not analytical on the null cone $U = 0$, which is the future light cone of the event $r=t=0$,
when the shockwave hits the massive particle. One can check that $H_{\mu\nu}^+(0,\Omega,p)=H_{\mu\nu}^-(0,\Omega,p)$, thus 
the amplitudes are continuous across $U=0$ but the derivative $\partial_U H_{\mu\nu}$ are not. As a result, some components of the Weyl tensor have a delta-function like singularities. We leave discussion of this property and its physical consequences for the next Section \ref{sec:3.4}. Note that, this non-analytical behavior holds for all $r$ and it is not related to the approximation method we use.

To see that (\ref{4.17a}), (\ref{4.17b})  have a form of outgoing gravitational radiation we analyze the structure of 
$H_{\mu\nu}$ in coordinates (\ref{4.11.1}). First we use constraints \eqref{4.9}, imposed by the gauge conditions, and definitions (\ref{4.17b})  to conclude that
\begin{equation} \label{5.1}
	\tilde{m}^\mu\left( \tilde{\beta}_{\mu\nu} - \frac{1}{2}\eta_{\mu\nu} \tilde{\beta} \right) = 0~,
\end{equation}
where  $\mathbf{\tilde{m}} = 2\tilde{n}_v \mathbf{l}(\tilde{n}) $, see \eqref{4.13}. One can check that $\tilde{m}_\mu = \delta^U_\mu$. Thus, $\mathbf{\tilde{m}} $ is a null normal vector to the light cone $U=0$. It follows from \eqref{4.15} and  \eqref{5.1} that
\begin{equation}\label{5.2}
	H_{r\nu} -\frac 12 \eta_{r\nu}H=0~,
\end{equation} 
\begin{equation}\label{5.3}
	H_{rr} =H_{rA}=\gamma^{AB}H_{AB}=0~.
\end{equation} 
Also there are residual coordinate transformations, discussed in Appendix \ref{App2},  which can be used to change $H_{U\mu}$ leaving 
angular components $H_{AB}$ invariant.
Thus, one  can impose additional conditions
\begin{equation}\label{5.8}
H_{U\mu}=0~.
\end{equation}
By taking into account \eqref{5.3} one can readily conclude that the only non-vanishing components are spherical traceless components
$H_{AB}$. At fixed $\Omega$ tensor $H_{AB}(U,\Omega,p)$ describes oscillations of the gravitational wave orthogonal to the wave propagation in the direction $\vec{e}(\Omega)$ . 

In the Bondi-Sachs formalism $H_{AB}$ is the strain tensor  (usually denoted as $C_{AB}$) and it describes 
two polarizations of outgoing gravitational waves, see, for example, \cite{Madler:2016xju}, while the time derivative
\begin{equation}\label{6.3}
N_{AB} = \frac{1}{2} \partial_U H_{AB}~,
\end{equation}
is called the Bondi news tensor. 

In the linearized approximation perturbations in collision of null and massive particles can be found from (\ref{4.15}) with the help of 
(\ref{4.1})  
\begin{equation}\label{4.15abc}
h_{\mu\nu}(x) =\frac{H_{\mu\nu}(U,\Omega)}{r}  + O(r^{-2})~,
\end{equation}
\begin{equation} \label{4.1b}
H_{\mu\nu}(U,\Omega) = \frac{1}{(2\pi)^2} \int d^2p~ \tilde{f}(p)~H_{\mu\nu}(U,\Omega,p) ~.
\end{equation}
The Fourier transform  $\tilde{f}(p)$ for the shockwave produced by the massless particle is given by (\ref{6.2}).

To calculate the flux of energy emitted in the collisions one needs the news tensor $N_{AB}(U,\Omega)=\frac{1}{2}\partial_U H_{AB}(U,\Omega)$.
In the Bondi-Sachs formalism the intensity of the energy flux at the future null infinity $\mathcal{J}^+$  (at $r\to \infty$)  in the 
direction $\vec{e}(\Omega)$ per unit area $d\Omega$ is
\begin{equation} \label{6.4}
I(U,\Omega) = \frac{1}{ \kappa}  N_{AB}(U,\Omega) N^{AB}(U,\Omega)~,
\end{equation}  
where indices are risen and lowered with the help of metric $\gamma_{AB}$ on the unit $S^2$.
Integrating over momenta $p$ in (\ref{4.1b}) one finds the intensity of the radiation energy flux at $U>0$
\begin{equation} \label{6.5}
I(U,\Omega)\Bigl|_{U>0} = \frac{2}{(1-e_x^2)^2 }   \left( (N_{\phi\phi})^2 + (1-e_x^2)(N_{\theta\phi})^2  \right)~,
\end{equation}
\begin{equation} \label{6.6}
H_{\theta\phi}(U,\Omega)  =  \frac{C}{\sqrt{1-e_x^2}}~ \mathrm{Re} \Biggl(\frac{i e(eU/a - 1+e_x^2  )}{    1-e_x -eU/a } 
	-\frac{2(be_2+e_1)}{c(1+b^2)}  \left(  be_1-e_2  + \frac{     1-e_x^2  }{ e_2 +ic}  \right) 
	\Biggr)~,
\end{equation}
\begin{multline} \label{6.7}
H_{\phi\phi}(U,\Omega)  = C~\mathrm{Re}   \Biggl(-\frac{ e(eU/a - (1-e_x)(1-e_x^2)/2  )}{     1-e_x -eU/a } 
	\\
	+\frac{1}{c}\left(   \frac{ 2(b e_2+e_1)^2}{ 1+b^2}- (1-e_x)(1-e_x^2) \left( \frac{1}{2} + 	\frac{  b e_1 - e_2 }{ (e_2 +ic)(1+b^2) } \right)
	\right)
	\Biggr)~,
\end{multline}
\begin{equation} \label{6.8}
C=	\frac{r_g \omega}{2a } ~,~b	=
	- \frac{U/a }{ e_2 +ic } ~,~ c^2=(U/a-e_1 )^2+  (1-e_x)^2 ~,~ e = e_1+ie_2~. 
\end{equation}
The expression for the flux at $U<0$ follows from (\ref{6.5})-(\ref{6.7}) on the base of the time-reversal symmetry
\begin{equation} \label{6.10}
I(U,\Omega)=I(-U,\Omega')~, 
\end{equation}
accompanied by the spatial reflection: 
\begin{equation} \label{6.11}
e_x(\Omega)=e_x(\Omega')~,~e_i(\Omega)=-e_i(\Omega')~,
\end{equation}
which corresponds to rotation by angle $\pi$ around  direction $e_x=1$.  This property is a consequence of
the symmetry properties
\begin{equation}\label{6.13}
N_{AB}(U,\Omega,p)=-N_{AB}(-U,\Omega',p)~,
\end{equation}
dictated by (\ref{4.17a}), (\ref{4.17b}) and by the relations:
\begin{equation}\label{6.12}
\tilde{k}_\pm(p,\Omega) = -\tilde{k}_\mp (p,\Omega')~,~ \tilde{\beta}_{AB}(p,\Omega)=(\tilde{\beta}_{AB})^*(p,\Omega')~,
\end{equation}
Time-reversal symmetry (\ref{6.10}) demonstrates that the total energies emitted in the collision before the massive particle meets the shockwave front and after coincide.  

Transformations $U\to -U$, $\Omega\to \Omega'$ act on the future null infinity $\mathcal{J}^+$. They have two fixed points:  $U=0$, $e_x=\pm 1$. Coordinates $U=0$, $e_x= 1$ mark position of a point on $\mathcal{J}^+$ which the null particle is asymptotically approaching. 

\begin{figure} 
	\centering
	\includegraphics[width=1\textwidth]{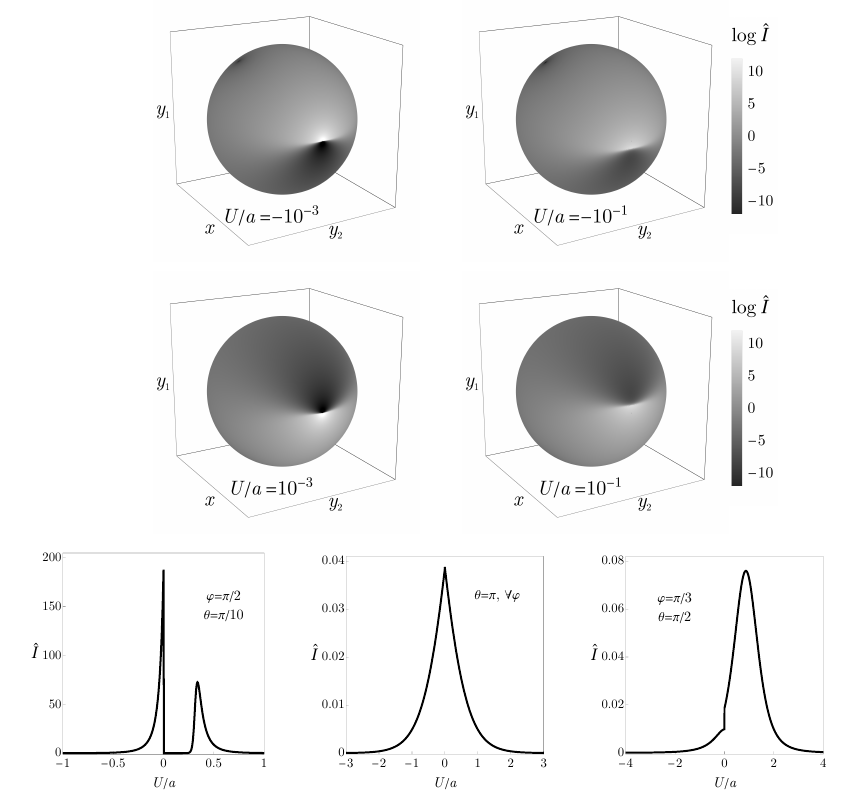}
	\caption{The angular distribution of the radiation energy flux $\hat{I}(U,\Omega) $ (in dimensionless units) in collision of null and massive particles.
		The null particle moves along the $x$-axis along the $y_1$-axis, the massive particle is located at the center of coordinates.
		  The distributions (top, $U<0$) and (middle, $U>0$) are shown for two moments of
		time $U$ on a logarithmic scale. Top figures demonstrate time-reversal symmetry of the flux.  Angle $\phi$ changes counterclockwise in the $y_1$-$y_2$ plane. The
		intensity of the flux grows in the direction of the null particle velocity as $\theta \to 0$.  The time evolutions are also shown for fixed
		spherical angles $\theta, \phi$ on bottom figures. The jump of the radiation energy flux  at $U=0$ is due to the presence of a spherical shockwave.}
	\label{fig}
\end{figure}

The results for the radiation energy flux  are presented on the figure \ref{fig} in dimensionless units $\hat{I}(U,\Omega) = (a^2/r_g \omega)^2 I(U,\Omega)$. Since the null particle  moves along the $x$-axis and is shifted along the $y_1$-axis to the lower hemisphere, the main peak (bright area) of the radiation at short positive $U$ is located in a narrow region near its trajectory, when $\theta \to 0$ and $\phi$ varies in the range $(\pi,2\pi)$.  
The duration of the peak is determined by the impact parameter $a$, it as about $0.1 a$. The height of the peak is proportional to $(r_g \omega/a^2)^2$.

When $e_x\to 1$ ($\theta \to 0$) and $U\to 0$ components $N_{AB}$, which follow from (\ref{6.6}), (\ref{6.7}), behave as $(Ua+c\theta)^{-2}$, where $c$ is a function of $\varphi$. As a result, the intensity $I(U,\Omega)$ is an analytic function on 
$\mathcal{J}^+$ except a pole at $U=\theta=0$.

The pole $U=\theta=0$ is a position of the null particle on $\mathcal{J}^+$, if the null particle is assumed to be a point-like.
It is reasonable to assume that our approximation for the gravitational field of the particle is restricted by values $\rho >EG$, where $EG$ is interpreted as an analogue of gravitational radius. Therefore particle's coordinates in the point-like limit should be determined with precision $O(EG)$. If trajectory of the null particle is determined with the precision $\delta u\simeq \delta y^i=O(EG)$ its location on
$\mathcal{J}^+$ cannot be more accurate than $|U|\sim U_0$, $\theta=0$, where $U_0=O(EG)$. (In fact, shifts $\delta x^\mu$ of coordinates in the bulk result in a BMS-like transformation $\delta U=\alpha(\Omega)$ of coordinates on $\mathcal{J}^+$, see \cite{Madler:2016xju}). 

The restriction $|U|\geq U_0$,  allows one to avoid the pole and consider $I(U,\Omega)$ at all $\Omega$. This corresponds to the peak intensity
\begin{equation}\label{8.10}
I_{\mbox{\tiny{peak}}}=I(U_0,\theta=0) =  \frac{C^2}{2\kappa}  \frac{a^2}{U_0^4} \sim  \frac{1}{\kappa} \left(  \frac{r_g}{GE} \right)^2~.
\end{equation}
Alternatively,  one can avoid the pole under the restriction $\theta \geq \theta_0$. One can check, with the help of  (\ref{8.10}), that $I(0,\theta_0)=I_{\mbox{\tiny{peak}}}$, if $\theta_0=O(EG/a)$.  Then the intensity can be considered at all $U$.

\subsection{Spherical shockwaves generated in fixed-target collisions}\label{sec:3.4}

The results of the previous Section demonstrate that observers at $\mathcal{J}^+$ register abrupt change of the intensity 
$[I](\Omega)$ of the gravitational radiation at the moment $U=0$. According to (\ref{6.4}), (\ref{6.13})
\begin{equation} \label{6.10a}
[I](\Omega)=I(0,\Omega)-I(0,\Omega')\neq 0~,~[N_{AB}](\Omega)=N_{AB}(0,\Omega)+N_{AB}(0,\Omega')\neq 0~.
\end{equation}
Thus, normal derivatives of the perturbations at $U=0$ are discontinuous on the null cone. This property can be interpreted as a spherical shockwave born at the moment when the massive particle meets the front of the initial plane shockwave.

Solution to the Einstein equations in the form of a spherical gravitational wave propagating in a flat spacetime 
is known as the Robinson-Trautman solution \cite{Robinson:1960}. Further studies of this solution can be found in \cite{Penrose:1972xrn}, \cite{Hogan:1993xj} and in other publications. In this work we consider the linearized gravity, so, before we proceed, some remarks on perturbations interpreted as spherical shockwaves are in order.

Note that perturbations $h_{\mu\nu}$ satisfy the wave equation (\ref{3.15}) and the gauge conditions (\ref{3.22}) at all $U$.
If one requires that $[h_{\mu\nu}]=0$ at $U$, Eqs. (\ref{3.15}), (\ref{3.22}) allow the following discontinuities in the normal derivatives
at $U=0$:
\begin{equation} \label{8.1}
[\partial_U	h_{UU}]=\frac{2[N_{UU}]}{r}~,~[\partial_U	h_{Ur}]=\frac{2[N_{Ur}]}{r}~,~
[\partial_U	h_{UA}]=2[N_{UA}]~,~
[\partial_U	h_{AB}]=2r[N_{AB}]~,
\end{equation}
\begin{equation} \label{8.2}
[\partial_U	h_{rr}]=[\partial_U	h_{rA}]=\gamma^{AB}[\partial_U	h_{AB}]=0~.
\end{equation}
Conditions (\ref{8.2}) are in accord with \eqref{5.3}.  Except these conditions quantities $[N_{\mu\nu}]=[N_{\mu\nu}](\Omega)$ can be arbitrary.  Note that four-velocities of geodesics which cross the light cone are continuous on $U=0$. However 
$[\partial_U	h_{\mu\nu}]$ in  (\ref{8.1}) result in discontinuities of the first derivatives of the velocities.

To see if those discontinuities require any null shell on the cone one can compute the transverse curvature
given by (\ref{2.11.1})
\begin{equation}\label{8.3}
\mathbb{C}_{ab}= e_a^\mu e_b^\nu \nabla_\nu n_\mu~,~U=0~,
\end{equation}
where the tetrade in the coordinates (\ref{4.11.1})  are defined as
\begin{equation}\label{8.4}
n=\frac{1}{\sqrt{2}}(2\partial_U-\partial_r)~,~~l=\frac{1}{\sqrt{2}}\partial_r~,~e_i=\frac{g_i}{r}\partial_i~,
\end{equation}
with $i=(\theta, \varphi)$ and $g_1=1$, $g_2=\sin^{-1}\theta$.
Vectors $e_a=(e_v,e_i)$ are tangent to the cone, $e_v=l$ is null normal to $U=0$. A straightforward computation with the help of 
(\ref{8.1}),(\ref{8.2})  yields
\begin{equation}\label{8.5}
[\mathbb{C}_{vv}]=[\mathbb{C}_{vi}]=0~~,~~[\mathbb{C}_{ij}]=\frac{2}{r}g_ig_j [N_{ij}]~~,~~[\mathbb{C}_{ii}]=\frac{2}{r}\gamma^{AB}[N_{AB}] =0~~.
\end{equation}
Thus, the considered spherical shockwave does not require any null matter propagating along the null cone $U=0$, the same property 
as that of the Robinson-Trautman solution. The two non-vanishing components of $[\mathbb{C}_{ab}]$, which can be joint in
the radiative strain
\begin{equation}\label{8.6}
\sigma_0(\Omega)=\frac{r}{2} ([\mathbb{C}_{11}]-i[\mathbb{C}_{12}]) =[N_{\theta\theta}] - \frac{i}{\sin\theta}[N_{\theta\phi}] ~,
\end{equation}
are related to two polarizations of the shockwave \cite{Barrabes:book}. The strain is determined by the discontinuity of the Bondi news tensor $[N_{AB}]$ on the null cone. The other components $[N_{U\mu}]$ are related to pure gauge variables, as has been explained above. This can be also seen by studying the curvature. In the retarded coordinates  in vicinity of the null cone one can write
\begin{equation}  \label{7.5}
h_{\mu\nu}(x) \bigl|_{U \to 0+} = h_{\mu\nu}^-(x) + U \Theta(U) [\partial_U h_{\mu\nu}(x)]\bigl|_{U=0}  + O(U)~.
\end{equation}
By using standard definitions for the Newman-Penrose invariants $\Psi_n$ in the basis $n,l,m=(e_1+ie_2)/\sqrt{2}$ one finds that non-analyticities (\ref{7.5}) appear only in $\Psi_3$ and $\Psi_4$. The scalar $\Psi_3=C_{\mu\nu\rho\sigma} l^\mu n^\nu \bar{m}^\rho n^\sigma$ has 
a jump on $U=0$ determined by $[N_{AB}]$  and its derivatives, while $\Psi_4$ acquires a singular contribution
\begin{equation}\label{8.8}
\Psi^{\mathrm{\tiny{sing}}}_4=C_{\mu\nu\rho\sigma} n^\mu \bar{m}^\nu n^\rho \bar{m}^\sigma= -\frac{2\sigma_0}{r} \delta(U)~.
\end{equation}
(Here the Weyl tensor $C_{\mu\nu\rho\sigma}$ is considered in the linear approximation.)

Explicit expressions for $[N_{AB}]$ on the null cone can be found from asymptotics (\ref{4.17a}), (\ref{4.17b})  which yield
\begin{equation}  \label{7.4}
	[\partial_U	h_{\mu\nu}(x,p)] = -\frac{\pi}{r\tilde{n}_v} \mathrm{Re}~ \tilde{\beta}_{\mu\nu}(x,p)~.
\end{equation}
Then from (\ref{4.1}) and (\ref{7.4}) by a straightforward calculation one gets 
the jump of the Bondi news tensor on the null cone $U=0$
\begin{equation}\label{8.9}
[N_{\theta\theta}] = \frac{\pi^2 r_g \omega}{a^2} \frac{(1+e_x^2)(-1+e_x^2+2e_1^2)}{(1+e_x)(1-e_x)^2}
	~,~
[N_{\theta\phi}]= \frac{4\pi^2 r_g \omega}{a^2} \frac{e_x e_1 e_2}{(1-e_x)\sqrt{1-e_x^2}}~
\end{equation}
in collision of null and massive particles. The radiative strain follows from (\ref{8.8}) and  definition (\ref{8.6}). Note that jumps (\ref{8.9})
diverge as $e_x\to 1$, as is expected.   The range of validity of (\ref{8.9}) can be restricted to satisfy the peak intensity (\ref{8.10}) at
$\theta \geq \theta_0$ with $\theta_0=O(EG/a)$.

\section{Discussion}\label{sec:4}

The main challenge of this work was to find a self-consistent formulation of the two-body problem, when a null particle moves in the gravitational field of a massive particle. Our aim was to find corrections to the gravitational field of the two bodies by solving the Einstein equations in the linearized approximation. 

The null particle produces a gravitational shockwave (GSW) whose wave front moves along a null hypersurface $\cal N$, a past event horizon of the particle. It has been known for a quite a long time \cite{tHooft:1985NPB,Sfetsos:1994xa,Blau:2015nee}  that in a flat spacetime and in some restricted class of geometries the GSW can be constructed by using supertranslations of $\cal N$, see \cite{Penrose:1972xrn}. The key condition of these works was that supertranslations have been considered as isometries of the space-time at $\cal N$. Given this condition  the GSW from the null particle yield no perturbations of the metric.  

The two-body problem addressed in this work is the illustration of property when supertranslations are violated on $\cal N$. As a result, the null particle produces perturbations in ${\cal M}^+$, the future to $\cal N$. Finding the perturbations requires a careful formulation of a characteristic Cauchy problem with the initial data on $\cal N$. In the linearized theory the Cauchy data can be determined and perturbations can be found in a useful integral form. 

The fixed target collisions of the two particles result in a gravitational radiation. The angular distribution of the flux at $\mathcal{J}^+$ 
can be found in an analytic form, see Eqs.(\ref{6.5})-(\ref{6.7}). On null cone $U=0$, the intensity of the flux changes the angular dependence from $I(U,\Omega)$ to $I(U,\Omega')$, as if observer's position rotates by angle $\pi$ around the axis along velocity of the null particle. The fact that 
the intensity is discontinuous on $U=0$ indicates a new physical feature: the massive particle creates a spherical gravitational shockwave when the particle crosses the wave front of the initial GSW. 

Spherical gravitational shockwaves in Minkowsky spacetime are known as Robinson-Trautman solutions. In the framework of the  linearized approximation we have found spherical gravitational shockwaves given by (\ref{8.1}), (\ref{8.9}) in a background field of a massive source. We expect that similar spherical gravitational shockwaves is a generic feature of null shells interacting with massive bodies.

This work has been focused on mathematical aspects in the gravitational two body problem when one of the bodies is null. It is clear that the reported results can be used in different applications. A variety of electromagnetic effects related to null objects in astrophysics 
\cite{Fursaev:2025did}, \cite{Fursaev:2023lxq} implies their analogs in the gravitational sector. Ultrarelativistic bodies, which appear in the Early Universe, can be a potential source of the gravitational radiation when they move near clumps of matter.  If kinks, loops and cusps of cosmic strings are considered as null particles, our results provide an alternative mechanism of gravitational waves from such defects \cite{Damour:2000wa, Damour:2001bk}. 

The formalism developed here can be used to describe gravitational perturbations in more complicated problems: with 
null shells instead of a null particle, or a matter distribution instead of a massive particle. 
The gyraton solutions \cite{Frolov:2005in} also allow one to consider null particles with spins by using the Penrose junction conditions \cite{Podolsky:2017pth}. Massive point-like source with an angular momentum can be also included in the two-body problem.

We are planning to return to some of these topics in forthcoming works.


\section{Acknowledgments}

The authors are grateful to Adeela Afzal, Evgeny Davydov, Irina Pirozhenko for stimulating discussions.

\newpage
\appendix

\section{Definitions and comments to Sections \ref{sec:2.3}, \ref{sec:2.4}}\label{App1}
\setcounter{equation}0

In this paper we use special coordinates $x^\mu=(u,v,y^i)$ to describe a null hypersurface $\cal N$ embedded in a space-time $\cal M$. 
We require that $u$ and $v$ are null coordinates and equation of $\cal N$ is just $u=0$. As is well-known, the normal vector to the null hypersurface is a generator of null geodesics which belong to $\cal N$. We choose coordinates where null geodesics on $\cal N$ are described 
by equations $x^\mu=x^\mu(\lambda)$ with:
\begin{equation}\label{a1.6}
v(\lambda)=\lambda~,~y^i=y_0^i~,~u=0~,
\end{equation}
where $y_0^i$ are some constraints.
It is convenient to define the null vectors
\begin{equation}\label{a1.7}
l^\mu=2 x^\mu_{~,\lambda}(\lambda)~,
\end{equation} 
tangent to the geodesics. Since the following relation is required to hold:
\begin{equation}\label{a1.8}
l_\mu(x)=\alpha(x)~ \partial_\mu u~,~u=0~,
\end{equation}
the metric components in the chosen coordinates have the properties
\begin{equation}\label{a1.9}
g^{uu}=g^{ui}=0~,~g_{vv}=g_{vi}=0~,~u=0~.
\end{equation}
These equations imply other relations on $\cal N$:
\begin{equation}\label{a1.10}
g^{uv}=g_{uv}^{-1}~,~\alpha=g_{uv}=0~,~g=-g_{uv}^2 \det g_{ij}~,~u=0~.
\end{equation}
As a result the induced metric on the null hypersurface takes on the expected form $g_{ij}dy^idy^j$.
Given (\ref{a1.9}) one can check that geodesic equations on $\cal N$ hold in the form:
\begin{equation}\label{a1.11}
\nabla_l l=\kappa l~,~\kappa = g^{uv}(2g_{uv,v}-g_{vv,u})~,~u=0~.
\end{equation}

We consider a linearized gravity theory, where $g_{\mu\nu}=\eta_{\mu\nu}+h_{\mu\nu}$.
In this theory the components of the Einstein tensor $G_{\mu\nu}$ are 
\begin{equation}\label{a1.1}
G_{\mu\nu}=\frac 12 (\partial_\alpha\partial_\mu \bar{h}^{\alpha}_{\nu} +\partial_\alpha\partial_\nu \bar{h}^{\alpha}_{\mu}-
\Box \bar{h}_{\mu\nu}-\eta_{\mu\nu} \partial^\alpha \partial^\beta \bar{h}_{\alpha\beta})~,
\end{equation}
where $\bar{h}_{\mu\nu}=h_{\mu\nu}-\frac 12 \eta_{\mu\nu} h$. (Note that our convention for the Riemann tensor is $R^\lambda_{~\mu\nu\rho}=\Gamma^\lambda_{\mu\rho,\nu}-....$ ).
In the chosen coordinates $(u,v,y^i)$ the component
\begin{equation}\label{a1.2}
G_{vv}=\frac 12 (- \partial_v^2h_{ii}+2\partial_i\partial_v h_{iv}-\partial_i^2 h_{vv})~,
\end{equation}
does not contain derivatives over $u$ (normal derivative). Thus, in the Einstein gravity the corresponding equation is the constraint on $\cal N$. One can also check that components $G_{vi}$, $G_{vu}$ contain first derivatives $\partial_u h_{\mu\nu}$, while $G_{ij}$, $G_{ui}$, $G_{uu}$ include second derivatives $\partial^2_u h_{\mu\nu}$. 

Our interest is on the case when ${\cal M}={\cal M}_+\cup {\cal M}_-$, where ${\cal M}_\pm$ are  joint along the common boundary $\cal N$.  In the chosen coordinates the components of $h_{ab}$ (where $a,b$ correspond to $v, y^i$) are continuous across $\cal N$, but not smooth: derivatives $\partial_u h_{ab}$ may have jump at $u=0$. As a result, some components $G_{\mu\nu}$ may have singularities on $\cal N$.  On general grounds, one can write
\begin{equation}\label{a1.3}
\partial_u h_{ab}(u, \mathbf{x}) \bigl|_{u \to 0} \simeq \partial_u h^{-}_{ab}(x) +\Theta(u) [\partial_u h_{ab}(\mathbf{x})] + O(u)~,~ \mathbf{x}=\{v,y^i\}~,
\end{equation}
\begin{equation}\label{a1.4}
h_{u\mu}(x) \bigl|_{u \to 0} = h_{u\mu}^-(x) + \Theta(u) [ h_{u\mu}(\mathbf{x})]  + O(u)~,
\end{equation}
where $ \Theta(u)$ is the step-function and notation $[A]$ is introduced after (\ref{2.6}).
Then substitution of (\ref{a1.3}), (\ref{a1.4}) into (\ref{a1.1}) yields relation (\ref{2.16}) of Section \ref{sec:2.3}
\begin{equation}\label{a1.5}
[G^{ab}] \simeq \delta(u) C^{ab}~,~u=0~,
\end{equation}
which holds up to terms which are not singular. Here we took into account that $G^{vv} = 4 G_{uu}$,  $G^{vi} = -2G_{ui}$, $G^{ij} = G_{ij}$.

In the present paper we use the de Donder gauge (\ref{3.13}). Conditions (\ref{a1.9})  require that
\begin{equation}\label{a1.12}
\hat{h}_{va}=0~,~u=0~.
\end{equation}
If $\hat{h}_{va}\neq 0$ one can perform the coordinate transform to $\hat{h}_{\mu\nu}'=\hat{h}_{\mu\nu}+\chi_{\mu,\nu}+\chi_{\nu,\mu}$ , where 
$\chi_u=0$ and $\chi_a$ can be found from the Cauchy problem:
\begin{equation}\label{a1.13}
\Box \chi_a=0~,
\end{equation}
\begin{equation}\label{a1.14}
\chi_{v,v}=-\frac 12 \hat{h}_{vv}~,~\chi_{i,v}= -\hat{h}_{vi}-\chi_{v,i}~,~u=0~.
\end{equation}
For perturbations (\ref{1.2}) quantities $\chi_{a,\mu}$ are well-defined. 

Our final comment is on the stress-energy tensor of a null particle with trajectory (\ref{a1.6}) on $\cal N$.
The action of the particle in a background metric has the standard form
\begin{equation}\label{a2.2}
I[x,g]=\int d\lambda~ \bar{E}(\lambda) ~x^\mu_{~,\lambda}(\lambda)g_{\mu\nu}(x(\lambda))x^\nu_{~,\lambda}(\lambda)~,
\end{equation}
where $\bar{E}(\lambda)$ is an energy parameter. Variation of the action over the metric yields the stress-energy tensor of the null particle:
\begin{equation}\label{a2.3}
T_{\mbox{\tiny{(n)}}}^{\mu\nu}(x)=\frac{2}{\sqrt{-g}}   \frac{\delta I[x,g]}{\delta g_{\mu\nu}(x)}  =
\frac{1}{2\sqrt{-g}}\int d\lambda~\delta^{(4)}(x-x(\lambda))~\bar{E}(\lambda)  l^\mu(\bar{x}) l^\nu(\bar{x})~,
\end{equation}
\begin{equation}\label{a2.4}
T_{\mbox{\tiny{(n)}}}^{\mu\nu}(x)=
\frac{1}{2\sqrt{-g}}\delta(u) \delta^{(2)}(y-y_0) \bar{E}(v)  l^\mu l^\nu~.
\end{equation}
One can check that (\ref{a2.4}) covariantly conserves, $\nabla_\mu T_{\mbox{\tiny{(n)}}}^{\mu\nu}=0$, provided that
\begin{equation}\label{a2.5}
\partial_l \bar{E}+\kappa \bar{E}=0 ~.
\end{equation}
In the linearized theory one gets from (\ref{a2.4}), (\ref{a2.5}) 
\begin{equation}\label{a2.6}
T_{\mbox{\tiny{(n)}}}^{\mu\nu}=\delta(u)\sigma(y,v) l^\mu l^\nu~, 
\end{equation}
\begin{equation}\label{a2.7}
\sigma(y,v)=E(v)\delta^{(2)}(y,y_0)~,
\end{equation}
\begin{equation}\label{a2.8}
E(v)=E (1+4h_{uv}(v,y_0))+O(h^2)~,
\end{equation}
where $E$ is a constant and $\delta^{(2)}(y,y_0)=\delta^{(2)}(y-y_0)\det^{-1/2} g_{ij}$. (Note that in the linear order $\det{g_{ij}}=1$, since $h_{ii}=0$). The parameter $E(v)$ can be interpreted as an energy of a null particle carried through a constant $v$ section of $\cal N$.
Expression  (\ref{a2.6})  is the generalization of (\ref{1.6b}) for the stress-energy tensor of a null particle in Minkowsky space-time. 

\section{Comments on asymptotic form of perturbations}\label{App2}
\setcounter{equation}0

Consider integral formula \eqref{4.7} for $h_{\mu\nu}(x,p)$. In retarded coordinates (\ref{4.11.1}) the argument of the $\Theta$-function in \eqref{4.7} takes the form:
\begin{equation}\label{4.12}
	(x\cdot  {\bf l}) = \frac{1}{2 n_v}\Bigl(U+r \bigl(1+(\vec{m}\cdot \vec{e}) \bigr)\Bigr)~,
\end{equation}
where the vector $\vec{e}$ is defined in (\ref{4.13}) and has the explicit form
\begin{equation}
	e_x = \cos\theta~,~
	e_1  = \sin\theta \sin\phi~,~
		e_2  = \sin\theta \cos\phi~, 
\end{equation}
 and 
\begin{equation} \label{4.13a}
 \vec{m}(\Omega')=(m_x, m_i)~,~m_x = n^2_v-n_i^2~,~ m_i=2 n_v n_i~.
 \end{equation}
The subsequent analysis of the derivation of the asymptotic form of perturbations generated by shockwaves with elementary profiles is based on the work \cite{Fursaev:2024czx}.

Due to the exponent in \eqref{4.7} the main contribution to the integral comes from a domain where the factor $\vec{m} \cdot \vec{e}+1$ is small, that is, $\vec{m}$ is almost $-\vec{e}$. It is this domain where the saddle-point approximation can be safely used.
Keeping this in mind, we split the integration in \eqref{4.7} into the region $\vec{m} \cdot \vec{e}+1 \leq \Lambda^2$ , and to the rest region, $\vec{m} \cdot \vec{e}+1 > \Lambda^2$, with $\Lambda$ being
a dimensionless parameter in the interval
\begin{equation} \label{4.14}
	\frac{|U|}{r} \ll \Lambda^2 \ll 1~.
\end{equation}
This yields 
\begin{equation}\label{4.15ab}
	h_{\mu\nu}(x,p) =r^{-1}\bigl(	H_{\mu\nu}(U,\Omega,p ) + \breve{H}_{\mu\nu}(\Lambda^2r,\Omega,p)\bigr)  + O(r^{-2})~,
\end{equation}
where $H_{\mu\nu}(U,\Omega,p )$ are defined by (\ref{4.17a}), (\ref{4.17b}). The static terms, which have been omitted in (\ref{4.15}),
are
\begin{equation}\label{4.18}
	\breve{H}_{\mu\nu}(\Lambda^2r,\Omega,p)=	-  
	\frac{ i \pi\tilde{\beta}_{\mu\nu} }{\tilde{k}_+}  \exp \left(  \frac{i \tilde{k}_+  r \Lambda^2}{2 \tilde{n}_v}  \right)~.
\end{equation}
The static terms depend on the cutoff and they are exponentially small at large $r$. These terms are not relevant for shockwaves from null particles (although in others cases \cite{Fursaev:2023oep} these terms may result in peculiar consequences). The amplitudes $H_{\mu\nu}$ are determined by the integration in the domain
$\vec{m} \cdot \vec{e}+1 \leq \Lambda^2$.  The method which yields (\ref{4.17a}), (\ref{4.17b})  requires 
that $e_x <1$.

Gauge conditions in \eqref{3.13} are invariant under residual coordinate transformations 
\begin{equation}\label{5.4}
	\delta x^\mu=\xi^\mu(x)~,~\Box \xi^\mu=0~.
\end{equation}
These transformations can be used to impose further restrictions on $H_{\mu\nu}$. We assume the following asymptotic  form in the Minkowsky coordinates
\begin{equation}\label{5.5}
	\xi_\mu(x,p)= r^{-1}(\sigma_\mu(U,\Omega,p)+\breve{\sigma}_\mu(\Omega,p))+O(r^{-2})~.
\end{equation}
One can show that equations $\Box \xi_\mu=0$  is satisfied in the leading order of $1/r$ expansion and subleading terms in \eqref{5.5} are determined by $\sigma_\mu$ and $\breve{\sigma}_\mu$.  Calculations also show that  \eqref{5.4} generate transformations of the metric perturbations $h'_{\mu\nu}=h_{\mu\nu}+\nabla_\mu \xi_\nu +\nabla_\nu \xi_\mu$ so that the components of the dynamical part now look as 
\begin{equation}\label{5.6}
	H'_{rr}=H_{rr}~,~H'_{rA}=H_{rA}~,~H'_{AB}=H_{AB}~,
\end{equation}
\begin{equation}\label{5.7}
	H'_{UU}=H_{UU}+2\partial_U\sigma_U~,~H'_{Ur}=H_{Ur}+\partial_U\sigma_r~,~
	H'_{UA}=H_{UA}+\partial_U\sigma_A~.
\end{equation}

\end{document}